\documentclass[conference]{IEEEtran}

\usepackage{tikz}
\usetikzlibrary{arrows.meta}
\usetikzlibrary{positioning}
\IEEEoverridecommandlockouts
\usepackage{cite}
\usepackage{amssymb,amsfonts}
\usepackage{algorithmic}
\usepackage{textcomp}
\usepackage{xcolor}
\usepackage{url}
\usepackage{multirow} 
\usepackage{multicol}
\usetikzlibrary{arrows.meta,positioning}
\usepackage[caption=false,font=footnotesize]{subfig}
\usepackage{amsmath}
\usepackage{booktabs}
\usepackage{enumitem} 
\usepackage{hyperref}
\usepackage[most]{tcolorbox}
\usepackage{graphicx}
\usepackage{siunitx} 
\usepackage[ruled,vlined]{algorithm2e}
\def\BibTeX{{\rm B\kern-.05em{\sc i\kern-.025em b}\kern-.08em
    T\kern-.1667em\lower.7ex\hbox{E}\kern-.125emX}}
    
\newcommand{\affmark}[1]{\textsuperscript{\normalfont #1}}
\newcommand{\corrmark}{\textsuperscript{\ensuremath{\dagger}}}

\begin{document}

\title{Explain, Then Forget: Recommendation Unlearning under Asymmetric Access Constraint}

\author{
    \IEEEauthorblockN{Haichao Zhang\affmark{1,2},
                      Chong Zhang\affmark{1,2},
                      Wen Wang\affmark{1},
                      Shi Qiu\affmark{1},
                      Peiyu Hu\affmark{1,2},
                      Jia Wang\affmark{1,2}\corrmark
                      }
    \IEEEauthorblockA{
        \affmark{1}Xi'an Jiaotong-Liverpool University
    }
    \IEEEauthorblockA{
        \affmark{2}University of Liverpool
    }
    
    \IEEEauthorblockA{
        \textbf{Email:}
        haichao.zhang22@student.xjtlu.edu.cn,
        jia.wang02@xjtlu.edu.cn
    }
    \thanks{\ensuremath{^\dagger}\,Corresponding author.}
    \thanks{This study was funded by the National Natural Science
    Foundation of China (Grant No.~72401232), the Natural Science
    Foundation of the Jiangsu Higher Educational Institution of
    China (Grant No.~23KJB520037), the XJTLU Research Development
    Fund (RDF-21-01-053), and Liaoning Liaohe Lab
    (Grant Nos.~LLL24ZZ-02-01 and LLL24ZZ-02-02).}
}

\maketitle

\begin{abstract}
Recommendation unlearning aims to remove the influence of specific historical interactions that should no longer affect future recommendations, such as noisy clicks or outdated user preferences. However, performing recommendation unlearning becomes much harder in modern \textbf{Machine-Learning-as-a-Service (MLaaS)} architectures, where the recommendation model is remotely hosted by a model provider (\textbf{MP}) and accessed by service operators (\textbf{SO}) only through APIs. In this setting, a service operator serves end users but cannot access or modify internal model parameters.
As a result, existing unlearning methods are difficult to apply because they require access to model internals for retraining, fine-tuning, or other parameter updates. To address these challenges, we propose \textbf{C}ausal \textbf{E}xplanation-based \textbf{U}nlearning ({CEU}), a novel unlearning framework tailored for MLaaS environments. Instead of modifying internal model weights, CEU operates through a two-step output-level score intervention. First, it estimates how the forgotten interaction causally contributes to the current recommendation. Second, guided by this causal signal, the system removes the influence of the forgotten interaction by adjusting the final ranking scores of affected items. Because CEU leaves model parameters untouched, it naturally supports continuous unlearning requests while avoiding the error accumulation that can result from repeated parameter updates. Extensive experiments on three public datasets show that CEU achieves unlearning effectiveness comparable to full retraining while preserving recommendation utility and significantly reducing computational cost. The source code is available at: \url{https://github.com/zhang-haichao/Explain-and-Forget}.
\end{abstract}

\section{Introduction}

Driven by growing model complexity and the high cost of training and maintenance, modern recommender systems increasingly adopt the Machine Learning as a Service (\textbf{MLaaS}) architecture~\cite{ribeiro2015mlaas,grigoriadis2023machine,philipp2020machine}. In this setting, the model provider (\textbf{MP}) trains and hosts the recommendation model, while the service operator (\textbf{SO}) serves end users by querying the model through APIs~\cite{ijcai2026_oneturnknockoutt}. This separation means that the SO can use the model but cannot access or modify its internal parameters. As the direct interface with users, the SO is responsible for maintaining the quality of recommendation results. However, user preferences change over time. Users may lose interest in previously preferred items or accidentally click irrelevant ones.
\begin{figure}[ht]
    \centering
    \vspace{-12pt}
    \begin{minipage}[c]{0.4\textwidth}
        \includegraphics[width=\textwidth]{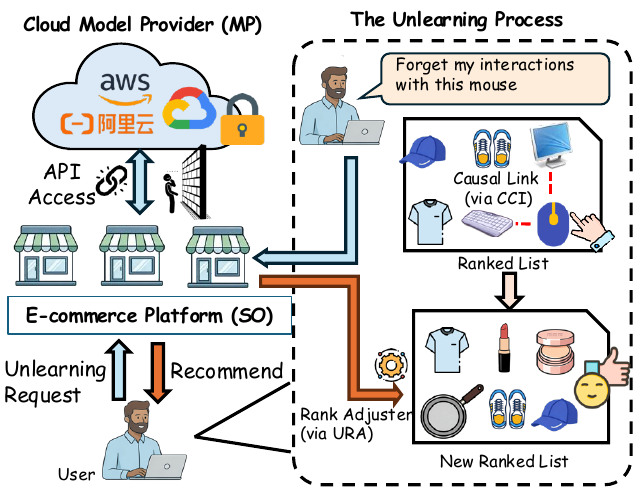}
    \end{minipage}\hfill
    \begin{minipage}[c]{0.48\textwidth}
    \caption{Illustration of recommendation unlearning in MLaaS. Since small E-commerce Platforms (SO) often lack the resources to train complex models, they rely on Cloud Model Providers (MP) to train models and serve recommendation results via APIs. To remove the influence of specific interactions, CEU uses CCI to identify causal contributions and uses URA to adjust rankings without modifying MP parameters.}
    \label{fig:ceu_framework}
    \end{minipage}
    \vspace{-5pt}
\end{figure}
Such noisy or outdated interactions can cause the system to repeatedly recommend unwanted content~\cite{sritrakool2021personalized,chen2019personal,hariri2015adapting}. 
To maintain user engagement, the SO should allow users to revoke specific historical interactions and remove their influence on future recommendations. This practical need motivates the task of recommendation unlearning~\cite{li2025survey}, which aims to remove the influence of selected historical interactions from a trained recommendation model.

Achieving recommendation unlearning in the MLaaS setting introduces two critical. First, a strict asymmetric access constraint: the SO has no access to the recommendation model's internal parameters or gradients. Since the model is controlled by the MP and exposed only through APIs, the SO cannot directly modify model weights to remove the effect of specific training data. Second, unlearning requests may arrive continuously as users interact with the system~\cite{lubitzsch2025towards,shen2025machine}. Repeated approximate updates across sequential requests may accumulate errors, gradually degrading overall recommendation quality~\cite{jiaqi2026harmonizing}.
These two constraints make existing unlearning methods difficult to apply, because most of them require access to model internals or control over the training process.
Existing methods can be broadly grouped into three categories: \textbf{retraining-based methods} (e.g., GraphEraser~\cite{chen2022graph}, RecEraser~\cite{chen2022recommendation}) that retrain affected model partitions; \textbf{fine-tuning-based methods} (e.g., GNNDelete~\cite{cheng2023gnndelete}, UnlearnRec~\cite{chen2025pre}) that apply lightweight representation updates; and \textbf{influence-based methods} (e.g., SCIF~\cite{li2023selective}, IFRU~\cite{zhang2024recommendation}) that adjust weights based on estimated data influence. Despite their different techniques, these approaches share a common limitation: they require access to model internals, training procedures, or updatable model components. As a result, they are difficult to apply in MLaaS systems where the model can only be accessed through APIs.

Given the strict constraints of the MLaaS paradigm, recommendation unlearning should shift away from parameter updates, as the Service Operator (SO) can only observe outputs via APIs. This suggests a different unlearning strategy: instead of updating model parameters, the system should revise recommendation outputs to remove the effect of forgotten interactions. However, executing output-level unlearning is non-trivial and presents two major challenges:
(1)\textbf{Which items to adjust?} To revise outputs correctly, the system must first identify which candidate items in the recommendation list were actually influenced by the forgotten interaction. (2)\textbf{How much to adjust?} Once the affected items are identified, the system must carefully penalize their scores to erase the forgotten data's influence. This adjustment must be highly precise. If the score reduction is too aggressive, it will distort the user's normal preferences; if it is too weak, the model will fail to truly \textit{forget} the revoked interaction, leaving unwanted data traces behind.

To address these challenges, we propose Causal Explanation-based Unlearning (\textbf{CEU}), a novel framework (illustrated in Fig.~\ref{fig:ceu_framework}) for recommendation unlearning through output-level score intervention in MLaaS environments, without modifying internal model parameters.
To address the first challenge, our framework employs a Causal Contribution Identifier (\textbf{CCI}). This module identifies the historical interactions that causally support a recommendation and estimates their contribution strengths.
Instead of exposing internal model parameters, the MP only needs to provide these attribution scores to the SO through the API interface. This design remains compatible with the MLaaS access setting without exposing model internals.
To address the second challenge, we introduce the Unlearning-driven Rank Adjuster (\textbf{URA}). Guided by the causal contributions provided by the CCI, the SO uses this module to adjust the scores of affected items and remove the influence of the forgotten interaction. Because CEU does not modify model parameters, it is naturally suited to continuous unlearning requests and avoids the error accumulation caused by repeated parameter updates. Appendix results under increasing forgetting ratios further support this property.
Our main contributions are summarized as follows:
\begin{itemize}
    \item We propose CEU, a novel recommendation unlearning framework designed for MLaaS environments. CEU removes the influence of forgotten interactions through output-level score intervention without modifying internal model parameters.
    \item We design the \textbf{Causal Contribution Identifier (CCI)} to identify the causal drivers of recommendations, thereby solving the attribution challenge in complex models.
    \item We introduce the \textbf{Unlearning-driven Rank Adjuster (URA)}, which removes the influence of forgotten interactions by applying localized score adjustments guided by the causal contributions identified by CCI.
    \item Extensive experiments on three public datasets show that CEU achieves unlearning effectiveness comparable to full retraining while preserving recommendation utility and significantly reducing computational cost.
\end{itemize}

\section{Related Work}

\subsection{Recommendation Unlearning}

Recommendation unlearning seeks to eliminate the influence of selected interactions from trained recommender models, e.g., removing outdated or noisy behaviors. Existing approaches can be broadly grouped into three categories. First, retraining-based methods (e.g., GraphEraser~\cite{chen2022graph}, RecEraser~\cite{chen2022recommendation}) aim for complete removal by partitioning data and retraining affected sub-models (SISA~\cite{bourtoule2021machine}). While exact, this paradigm severely limits efficiency and scalability. Second, fine-tuning-based methods (e.g., GNNDelete~\cite{cheng2023gnndelete}, UnlearnRec~\cite{chen2025pre}) avoid full retraining by applying lightweight representation updates to correct the influence of the forgotten data. Third, influence-based methods (e.g., SCIF~\cite{li2023selective}, IFRU~\cite{zhang2024recommendation}) efficiently approximate the gradient or loss impact of deleted data to adjust model weights. Furthermore, prompt-tuning-based recommendation unlearning has also been explored, where lightweight prompts and a teacher--student objective are used to forget target interactions while preserving useful retained knowledge (e.g., Prompt-Tuning for Recommendation Unlearning~\cite{11391920}). Despite their efficiency, both fine-tuning and influence-based approximate methods often suffer from incomplete forgetting, unintended performance drift, and accumulated errors during continuous sequential requests. Crucially, because all three paradigms require accessing model internals to update parameters, they are incompatible with the strict asymmetric access constraints of MLaaS. This motivates our parameter-free framework.

\subsection{Causal Explainable Recommendation}
Recent studies use causal and counterfactual perspectives to identify what truly drives a recommendation, generally falling into two directions. 
The first builds causality-aware models incorporating explicit reasoning structures. For example, using reinforcement learning on knowledge graphs~\cite{xian2019reinforcement} or neural symbolic reasoning (CAFE~\cite{xian2020cafe}) to trace underlying user intents.
The second focuses on post-hoc counterfactual explanations, analyzing trained black-box recommenders by asking ``what-if'' questions. CountER~\cite{tan2021counterfactual} searches for minimal input changes reversing a decision, while CF$^2$~\cite{tan2022learning} combines factual and counterfactual reasoning to capture necessity and sufficiency.
While these methods quantify causal influence for interpretation and transparency, they do not leverage these attributions to execute recommendation unlearning, especially under the strict constraints of MLaaS architectures.

\section{Preliminaries}
\subsection{Recommendation Task}
Let $\mathcal{U} = \{u_1, u_2, ...\}$ denote the set of users and $\mathcal{I} = \{i_1, i_2, ...\}$ denote the set of items. The historical user-item interactions are represented by the dataset $\mathcal{D} = \{(u, i) \mid u \in \mathcal{U}, i \in \mathcal{I}, R_{ui} = 1\}$, where $R_{ui}=1$ indicates an observed interaction. The training set is denoted as $\mathcal{D}_{\text{tr}} \subseteq \mathcal{D}$.

The primary goal in recommendation is to train a model $f$ with parameters $\theta$ using a learning algorithm $A$, such that $\theta = A(\mathcal{D}_{\text{tr}})$. This model $f$ is used to predict preference scores $\hat{y}_{ui} = f(u, i \mid \theta)$ for unobserved user-item pairs, from which a personalized top-K ranking list is generated for each user.

\subsection{Recommendation Unlearning}

Recommendation unlearning aims to efficiently eliminate the influence of specific data from a pre-trained model $f$.
Formally, let $\mathcal{D}_{\text{tr}}$ be the original training set. We define a "forgotten set" $\mathcal{D}_f \subseteq \mathcal{D}_{\text{tr}}$ as the collection of user-item interactions that must be removed. Correspondingly, the "retained set" is defined as $\mathcal{D}_r = \mathcal{D}_{\text{tr}} \setminus \mathcal{D}_f$.

The ideal but computationally prohibitive solution is to obtain a "gold model," $f_r$, by retraining from scratch using only the retained data. The parameters of this model are denoted as $\theta_r = A(\mathcal{D}_r)$.
Given the infeasibility of retraining, the practical task is to design an efficient unlearning method that approximates the recommendation behavior of this gold model. 
Parameter-based approaches achieve this by updating model parameters to obtain an unlearned model $f_u$ with weights $\theta_u$. In contrast, parameter-free methods like CEU perform unlearning through output-level adjustment without modifying the original $\theta$.

The objective is to design $U$ such that $f_u$ closely approximates the gold model $f_r$, while balancing three primary goals:
\begin{itemize}
    \item \textbf{G1: Unlearning Completeness.} The unlearned model $f_u$ must completely eliminate the influence of the forgotten data $\mathcal{D}_f$.
    \item \textbf{G2: Unlearning Efficiency.} The unlearning algorithm $U$ must be time-efficient, avoiding the high computational cost.
    \item \textbf{G3: Model Utility.} The unlearned model $f_u$ must maintain recommendation performance comparable to the gold model $f_r$.
\end{itemize}

\subsection{Unlearning as Causal Intervention}
\label{sec:unlearning_as_intervention}
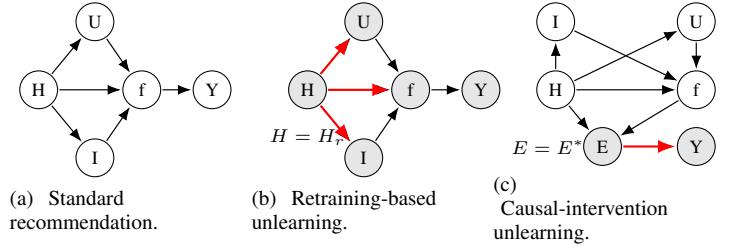
\begin{figure}[ht]
\centering

\captionsetup[subfloat]{
  justification=raggedright,
  singlelinecheck=false
}

\tikzset{
  >=Latex,
  every node/.style={font=\scriptsize},
  var/.style={
    circle,
    draw,
    minimum size=5mm,
    inner sep=0pt,
    fill=white
  },
  vgray/.style={
    circle,
    draw,
    minimum size=5mm,
    inner sep=0pt,
    fill=gray!20
  },
  edge/.style={
    -{Latex},
    line width=0.45pt,
    shorten >=0.5pt,
    shorten <=0.5pt
  },
  rededge/.style={
    -{Latex},
    line width=0.8pt,
    red,
    shorten >=0.5pt,
    shorten <=0.5pt
  }
}

\subfloat[
  Standard recommendation.
  \label{fig:causal_traditional}
]{
  \begin{minipage}[t]{0.25\linewidth}
  \vspace{0pt}
  \centering
  \begin{tikzpicture}[x=6.2mm,y=7.5mm]
    \node[var] (H) at (0,0) {H};
    \node[var] (U) at (1.2,1.2) {U};
    \node[var] (I) at (1.2,-1.2) {I};
    \node[var] (f) at (2.2,0) {f};
    \node[var] (Y) at (3.7,0) {Y};

    \draw[edge] (H) -- (U);
    \draw[edge] (H) -- (I);
    \draw[edge] (H) -- (f);
    \draw[edge] (U) -- (f);
    \draw[edge] (I) -- (f);
    \draw[edge] (f) -- (Y);
  \end{tikzpicture}
  \end{minipage}
}
\hfill
\subfloat[
  Retraining-based unlearning.
  \label{fig:causal_retraining}
]{
  \begin{minipage}[t]{0.25\linewidth}
  \vspace{0pt}
  \centering
  \begin{tikzpicture}[x=6.2mm,y=7.5mm]
    \node[vgray] (H) at (0,0) {H};
    \node[vgray] (U) at (1.2,1.2) {U};
    \node[vgray] (I) at (1.2,-1.2) {I};
    \node[vgray] (f) at (2.2,0) {f};
    \node[vgray] (Y) at (3.7,0) {Y};

    \draw[edge] (H) -- (U);
    \draw[edge] (H) -- (I);
    \draw[edge] (H) -- (f);
    \draw[edge] (U) -- (f);
    \draw[edge] (I) -- (f);
    \draw[edge] (f) -- (Y);

    \draw[rededge] (H) -- (U);
    \draw[rededge] (H) -- (I);
    \draw[rededge] (H) -- (f);

    \node[below=0.4em of H] {$H=H_r$};
  \end{tikzpicture}
  \end{minipage}
}
\hfill
\subfloat[
  Causal-intervention unlearning.
  \label{fig:causal_intervention}
]{
  \begin{minipage}[t]{0.25\linewidth}
  \vspace{0pt}
  \centering
  \begin{tikzpicture}[x=6.2mm,y=7.5mm]
    \node[var]   (H) at (0,0) {H};
    \node[var]   (U) at (3,1.2) {U};
    \node[var]   (I) at (0,1.2) {I};
    \node[var]   (f) at (3,0) {f};
    \node[vgray] (E) at (1,-1) {E};
    \node[vgray] (Y) at (3,-1) {Y};

    \draw[edge] (H) -- (U);
    \draw[edge] (H) -- (I);
    \draw[edge] (H) -- (f);
    \draw[edge] (U) -- (f);
    \draw[edge] (I) -- (f);
    \draw[edge] (H) -- (E);
    \draw[edge] (f) -- (E);
    \draw[rededge] (E) -- (Y);

    \node[below=0.8em of H, left=-0.4em of E] {$E=E^*$};
  \end{tikzpicture}
  \end{minipage}
}
\caption{
  Causal graphs of (a) standard recommendation,
  (b) retraining-based unlearning via $do(H:=H_r)$, and
  (c) CEU, which intervenes on $E\!\rightarrow\!Y$ to suppress
  forgotten-data effects without updating $f$.
  $H$, $U$, $I$, $f$, $E$, and $Y$ denote history, users, items,
  model, explanation, and output, respectively.
}
\label{fig:causal_graph}
\end{figure}

We use a Structural Causal Model (SCM) framework~\cite{pearl2010causal} as a conceptual view of recommendation unlearning and to motivate our intervention design.
Fig.~\ref{fig:causal_graph} (a) illustrates the standard recommendation SCM, where historical interactions ($H$) causally influence the recommendation output ($Y$) through multiple paths. These include a "front-door" path $H \to f \to Y$, as well as "back-door" paths mediated by user and item representations, $H \to U \to f \to Y$ and $H \to I \to f \to Y$.

Fig.~\ref{fig:causal_graph} (b) illustrates retraining-based unlearning as a global intervention on the node $H$, which can be described by the do-operator as:
$$
Y_r = P(Y \mid do(H := H_r)),
$$
where $H_r$ denotes the remaining historical interactions after removing the forgotten subset $H_f$ from $H$.
This intervention (red arrows) invalidates all downstream descendants (gray nodes $U, I, f, Y$), forcing their complete recomputation. While this "full-graph" intervention is causally complete (satisfying G1 and G3), it requires high-cost recomputation, thus violating G2.

As shown in Fig.~\ref{fig:causal_graph} (c), our approach approximates unlearning through a localized intervention guided by causal explanation. We augment the SCM by introducing an explanation node:  
\[
E = \text{explain}(H, f),
\] which represents the counterfactual contribution of $H$ to $Y$. 
This redirects the causal source of $Y$: 
we \emph{prune} the original causal edge $f \!\to\! Y$ 
and \emph{graft} a new path $E \!\to\! Y$ (red arrow). 
Without modifying any model parameters, we perform a localized intervention along this path:
\[
Y_r = P(Y \mid \text{do}(E := E^*)),
\]
where $E^*$ denotes the updated explanatory representation after removing the forgotten interaction.
This design is causally motivated rather than fully identified in the formal sense. It treats the explanation path as a practical approximation of how forgotten interactions affect the output and intervenes only in the part of the ranking process most directly linked to the forgotten effect, without changing global model parameters. 
In this way, it provides an efficient and interpretable approximation to recommendation unlearning, with the goal of balancing G1, G2, and G3 in practice.

\section{Proposed Methods}
\subsection{Overview}

\begin{figure*}[ht]
    \centering
    \vspace{-15pt}
    \includegraphics[width=0.74\linewidth]{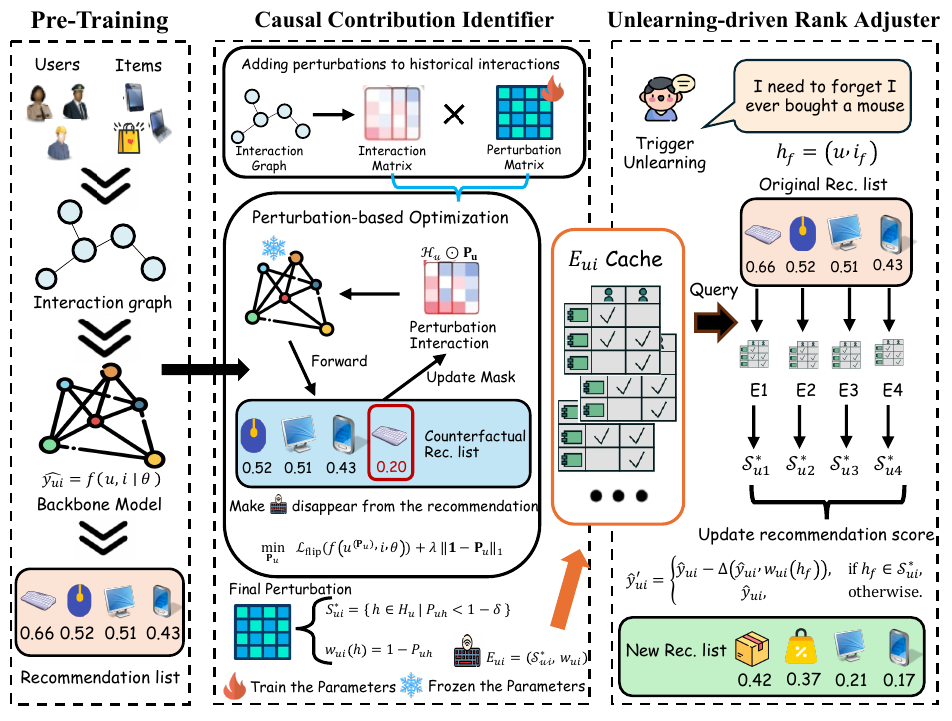}
    \caption{The framework of CEU. The \textbf{Pre-Training} pipeline generates the initial ranking results. The \textbf{Causal Contribution Identifier} identifies the minimal causal support set and quantifies their contribution to the recommendation. The \textbf{Unlearning-driven Rank Adjuster} performs localized score adjustment to remove the influence of the forgotten interaction.}
    \label{fig:framework_overview}
    \vspace{-12pt}
\end{figure*}

We propose Causal Explanation-based Unlearning (CEU), a unified framework that performs recommendation unlearning through a causal “explain-then-forget” paradigm.
The central idea is that unlearning should not rely on direct parameter modification but instead be guided by causal reasoning: before removing the influence of specific data, the system must understand why that data affects the recommendation outcome.
As illustrated in Fig.~\ref{fig:framework_overview}, CEU comprises two sequential modules:

\begin{itemize}
    \item \textbf{Causal Contribution Identifier (CCI) --- The Explain Stage.}
    In our MLaaS setting, the Cloud Model Provider (MP) hosts the CCI module. Since the MP has full access to the model, it directly computes the causal influence of past interactions and shares these attributions with the SO via a dedicated API.
    Specifically, for a user $u$ and a recommended item $i$, the CCI analyzes the user's interaction history $H_u$. It first identifies the Minimal Causal Support Set ($S_{ui} \subseteq H_u$), which contains the key interactions that primarily drive the recommendation. Then, it calculates the causal contribution strength, $w_{ui}(h)$, for each historical item $h \in H_u$.
    \item \textbf{Unlearning-driven Rank Adjuster (URA) --- The Forget Stage.} 
    When requested to unlearn a specific interaction $h_f = (u, i_f)$, URA uses the causal contribution weights provided by CCI. It performs a targeted intervention at the post-hoc ranking stage by adjusting the scores of candidate items. This precisely removes the influence attributed to $i_f$ along the identified causal path ($E \to Y$ in Fig.~\ref{fig:causal_graph}(c)), thereby mitigating side effects and achieving high efficiency and locality.
\end{itemize}

\subsection{Causal Contribution Identifier}
\label{sec:cci}
The Causal Contribution Identifier (CCI) module serves as the \textit{Explain} component within the CEU framework. It aims to uncover why a specific item $i$ is recommended to user $u$ by identifying the minimal set of historical interactions whose removal would reverse the recommendation decision.
\subsubsection{Explanation of recommendation}
Inspired by counterfactual reasoning~\cite{tan2021counterfactual}, CCI formulates the explanation generation process as searching for the smallest perturbation to the user’s history that flips the model’s output.
Let the historical interactions for user $u$ be denoted as: 
\[
\mathcal{H}_u = \{ (u, i_j) \mid (u, i_j) \in \mathcal{D}_{\text{tr}} \}.
\]

We define $\mathcal{S}_{ui}^{*}$ as the Causal Support Set, representing the minimal subset of the user’s historical interactions that are necessary and sufficient to cause the recommendation of item $i$.
\[
\mathcal{S}_{ui}^{*} = \operatorname*{arg\,min}_{\mathcal{S}\subseteq\mathcal{H}_{u}} |\mathcal{S}| \quad \text{s.t.} \quad \|f(u,i|\mathcal{H}_{u},\theta)-f(u,i|\mathcal{H}_{u}\backslash\mathcal{S},\theta)\| > \epsilon
\]
where $\epsilon$ is a threshold defining causal sensitivity.
We define the explanation for recommendation $(u, i)$ as a tuple:
\[
E_{ui} = (\mathcal{S}_{ui}^{*}, w_{ui}),
\]
where $\mathcal{S}_{ui}^{*}$ is the Causal Support Set, and $w_{ui}(h)$ is the \textit{Contribution Function} measuring the causal strength of each historical interaction $h \in \mathcal{S}_{ui}^{*}$. 
This formulation captures the counterfactual logic: \textit{``If the interactions in $\mathcal{S}_{ui}^{*}$ had not occurred, item $i$ would not have been recommended to user $u$.''}

\subsubsection{Quantifying Causal Contributions}
We obtain the explanation $E_{ui}$ by learning a minimal perturbation on user $u$’s history that reverses the recommendation of item $i$.  
We introduce a continuous mask vector $\mathbf{P}_u \in [0,1]^{|\mathcal{H}_u|}$ applied to the user’s interaction history $\mathcal{H}_u$, where each element $P_{uh}$ corresponds to one specific interaction $h \in \mathcal{H}_u$. 
The value $P_{uh}$ controls the retention degree of interaction $h$: a value of $1$ means the interaction is fully preserved, while smaller values indicate the interaction is suppressed or removed. 
Thus, $\mathbf{P}_u$ defines a weighted version of the user history, the masked history is $\mathcal{H}_u \odot \mathbf{P}_u$.

Following the Occam’s Razor principle~\cite{rasmussen2000occam}, we aim for explanations that are both simple and effective.  
To obtain such an explanation, we solve the following optimization problem:
\[
\min_{\mathbf{P}_u}\;
\mathcal{L}_{\mathrm{flip}}\!\big(f(u^{(\mathbf{P}_u)}, i \mid \theta)\big)
+\lambda \,\|\mathbf{1}-\mathbf{P}_u\|_{1},
\]
where $u^{(\mathbf{P}_u)}$ denotes the counterfactual user obtained by weighting the historical interactions of user $u$ with the mask vector $\mathbf{P}_u$, and $\mathcal{L}_{\mathrm{flip}}$ enforces a significant decrease in the predicted score for item $i$, and the $\ell_{1}$ term encourages a sparse perturbation (few interactions altered). 
A practical choice for the flip loss uses a hinge-margin against the $(K\!+\!1)$-th item $i_{K+1}$ in $u$'s original list, where $K$ is the cutoff of the top-$K$ recommendation list for user $u$, and $i_{K+1}$ denotes the item ranked $(K+1)$-th in the original ranking.
\[
\mathcal{L}_{\mathrm{flip}} = \max\big(0,\, m + f(u^{(\mathbf{P}_u)}, i \mid \theta) - f(u,i_{K+1} \mid \theta)\big),
\]
with margin $m > 0$. 
Since the MP executes the CCI module, it has full access to the internal parameters $\theta$. Therefore, the MP can directly optimize the mask $P_u$ using standard gradient-based updates, while keeping the model weights strictly frozen. 
After optimization, interactions with $P_{uh} < 1-\delta$ (a small threshold $\delta$) are deemed causal. 
The resulting \emph{Causal Support Set} is
\[
\mathcal{S}_{ui}^{*} = \{\, h \in \mathcal{H}_u \;|\; P_{uh} < 1-\delta \,\},
\]
which represents the minimal subset of the user’s historical interactions that are necessary and sufficient to cause the recommendation of item $i$. 
We define the per-interaction \emph{causal contribution weight} as
\[
w_{ui}(h) = 1 - P_{uh},
\]
which quantifies the strength of $h$ in sustaining $\hat{y}_{ui} = f(u,i \mid \theta)$. 
The explanation for $(u,i)$ is $E_{ui}=(\mathcal{S}_{ui}^{*},\, w_{ui})$, which will be used by the unlearning module.

\subsection{Unlearning-driven Rank Adjuster}
\label{sec:ura}
The Unlearning-driven Rank Adjuster (URA) module executes the \textit{Forget} phase of the CEU framework.
It performs efficient and precise recommendation unlearning based on the causal attributions generated by the CCI module (Section~\ref{sec:cci}). 
Its objective is to remove the influence of a forgotten interaction $(u, h_f)$ without retraining or altering model parameters. 
URA achieves this by locally adjusting the ranking scores of items affected by the forgotten cause.

\subsubsection{Causal-Guided Unlearning}

When a forgetting request is received for a specific user–interaction pair $h_f = (u, i_f)$, URA first identifies the potential propagation of its causal effect.  
For each candidate item $i$ in the user’s current recommendation list $L_u = \{ i_1, i_2, \dots, i_K \}$, the SO queries the MP's attribution API to retrieve the causal explanations $E_{ui}=(\mathcal{S}_{ui}^*, w_{ui})$. This lightweight interaction allows the SO to obtain the necessary causal guidance without accessing any internal model parameters.
Here, $\mathcal{S}_{ui}^*$ is the Causal Support Set describing which historical interactions of user $u$ are causally responsible for recommending $i$, and $w_{ui}(h)$ quantifies the causal contribution of each interaction $h \in \mathcal{S}_{ui}^*$.  
These explanations can be precomputed and cached for real-time unlearning.
\subsubsection{Localized Score Adjustment}
URA adjusts the predicted preference scores based on the causal relevance of the forgotten interaction.  
Let $\hat{y}_{ui} = f(u, i \mid \theta)$ denote the original prediction for user $u$ and item $i$.  
The adjusted score $\hat{y}'_{ui}$ is computed as:
\[
\hat{y}'_{ui} =
\begin{cases}
\hat{y}_{ui} - \Delta(\hat{y}_{ui}, w_{ui}(h_f)), & \text{if } h_f \in \mathcal{S}_{ui}^{*},\\[3pt]
\hat{y}_{ui}, & \text{otherwise,}
\end{cases}
\]
where only items whose recommendations are causally dependent on $h_f$ are modified, ensuring that interventions remain local and precise.  

The penalty function $\Delta(\cdot,\cdot)$ determines the degree of attenuation according to the causal contribution $w_{ui}(h_f)$.  
A simple yet effective form is the multiplicative penalty:
\[
\Delta(\hat{y}_{ui}, w_{ui}(h_f)) = \alpha \cdot \hat{y}_{ui} \cdot w_{ui}(h_f),
\]
which proportionally reduces $\hat{y}_{ui}$ based on the strength of causal dependence.  
A higher $w_{ui}(h_f)$ leads to a larger decrease in $\hat{y}_{ui}$, reflecting stronger causal influence.

\subsubsection{Reconstruction of the Recommendation Ranking}

After computing the adjusted scores $\{\hat{y}'_{ui} \mid i \in L_u\}$ for all candidate items, URA reconstructs the user’s recommendation list by re-ranking items according to $\{\hat{y}'_{ui}\}_{i \in L_u}$.  
The final unlearned ranking is denoted as:
\[
L'_u = \text{Rank}\big(\{\hat{y}'_{ui}\}_{i \in L_u}\big).
\]

This operation effectively removes the causal effect of $h_f$ from user $u$’s recommendations while preserving unaffected preferences.
Because CEU does not update model parameters, it naturally avoids the error accumulation that may arise in sequential unlearning requests. Additional results under increasing unlearning ratios are reported in the Appendix.

\begin{table*}[ht]
\centering
\vspace{-15pt}
\caption{Recommendation utility after unlearning under a 1\% forgetting ratio. Origin and Retrain are reference results only and are excluded from method comparison; bold denotes the best practical unlearning method. Results are averaged over 3 runs (p $<$ 0.05, paired t-test).}
\label{tab:RQ1}
\vspace{-8pt}
\begin{tabular}{cccccccccc}
\toprule
\multirow{2}{*}{\textbf{ML-1M}} & \multicolumn{4}{c}{\textbf{BPR}} & \multicolumn{4}{c}{\textbf{LightGCN}} \\
\cmidrule(lr){2-5} \cmidrule(lr){6-9}
 & \textbf{HR@10} & \textbf{NDCG@10} & \textbf{HR@20} & \textbf{NDCG@20} & \textbf{HR@10} & \textbf{NDCG@10} & \textbf{HR@20} & \textbf{NDCG@20} \\
\midrule
Origin & 0.7520 & 0.2526 & 0.8444 & 0.2597 &  0.7680 & 0.2715 & 0.8637 & 0.2787 \\
Retrain & 0.7419 & 0.2417 & 0.8488 & 0.2525 &  0.7649 & 0.2672 & 0.8613 & 0.2696 \\
\hline
GraphEraser & 0.4540 & 0.0902 & 0.6126 & 0.0995 & 0.5635 & 0.1295 & 0.6968 & 0.1388 \\
RecEraser & 0.4833 & 0.1055 & 0.6260 & 0.1123 & 0.6003 & 0.1467 & 0.7379 & 0.1574 \\
SCIF & 0.5341 & 0.1332 & 0.6833 & 0.1436 & 0.6060 & 0.1723 & 0.7230 & 0.1764 \\
IFRU & 0.4227 & 0.2248 & 0.5749 & \textbf{0.2634} & 0.4350 & 0.2205 & 0.6502 & \textbf{0.2746} \\
GNNDelete & 0.6088 & 0.1212 & 0.7897 & 0.1511 & 0.5139 & 0.1305 & 0.6375 & 0.1390  \\
UnlearnRec & 0.5784 & 0.2206 & 0.7396 & 0.2318 & 0.4937 & 0.1320 & 0.6219 & 0.1415 \\
\textbf{CEU} & \textbf{0.7373} & \textbf{0.2533} & \textbf{0.8364} & 0.2602 & \textbf{0.7584} & \textbf{0.2671} & \textbf{0.8535} & 0.2736 \\
\toprule
\multirow{2}{*}{\textbf{Netflix}} & \multicolumn{4}{c}{\textbf{BPR}} & \multicolumn{4}{c}{\textbf{LightGCN}} \\
\cmidrule(lr){2-5} \cmidrule(lr){6-9}
 & \textbf{HR@10} & \textbf{NDCG@10} & \textbf{HR@20} & \textbf{NDCG@20} & \textbf{HR@10} & \textbf{NDCG@10} & \textbf{HR@20} & \textbf{NDCG@20} \\
\midrule
Origin & 0.8550 & 0.2365 & 0.9453 & 0.2024 & 0.8647 & 0.2418 & 0.9497 & 0.2043 \\
Retrain & 0.7805 & 0.1631 & 0.9051 & 0.1750 & 0.8122 & 0.2020 & 0.9226 &  0.1930 \\
\hline
GraphEraser & 0.6272 & 0.1005 & 0.8199 & 0.0938 & 0.7118 & 0.1225 & 0.8563 & 0.1165 \\
RecEraser & 0.6795 & 0.1079 & 0.8225 & 0.1021 & 0.7228 & 0.1319 & 0.8880 & 0.1266 \\
SCIF & 0.4020 & 0.0748 & 0.5239 & 0.0829 & 0.4825 & 0.0935 & 0.6377 & 0.1045 \\
IFRU & 0.2037 & 0.0938 & 0.2997 & 0.1192 & 0.2556 & 0.1373 & 0.3452 & 0.1744 \\
GNNDelete & 0.7515 & 0.1622 & 0.8879 & 0.1703 & 0.5473 & 0.1690 & 0.6932 & 0.1785   \\
UnlearnRec & 0.7267 & 0.1641 & 0.8667 & 0.1751 & 0.6318 & 0.1670 & 0.7192 & 0.1715 \\
\textbf{CEU} & \textbf{0.8527} & \textbf{0.2343} & \textbf{0.9437} & \textbf{0.2006} & \textbf{0.8627} & \textbf{0.2402} & \textbf{0.9480} & \textbf{0.2031} \\
\bottomrule
\vspace{-13pt}
\end{tabular}
\end{table*}

\section{Experiments}

We conduct extensive experiments on three public datasets to evaluate the proposed CEU framework, aiming to answer the following four research questions:
\begin{itemize}
    \item \textbf{RQ1: Utility Preservation (G3).}  
    Can CEU preserve recommendation utility on the retained data while enforcing unlearning, while remaining close to the original model and the retrained reference?
    
    \item \textbf{RQ2: Unlearning Effectiveness (G1).}  
    Does CEU successfully eliminate the influence of forgotten interactions from future recommendations?
    
    \item \textbf{RQ3: Time Efficiency (G2).}  
    What is the end-to-end latency and computational cost of CEU under both single and batched unlearning requests?
    
    \item \textbf{RQ4: Ablation on CCI.}  
    How crucial is the causal attributions step (CCI) to the overall effectiveness of CEU?
\end{itemize}

\subsection{Experiment Setup}
\subsubsection{Datasets}
We evaluate CEU on three public benchmarks (Table~\ref{tab:dataset_stats}): ML-100K\footnote{\url{https://grouplens.org/datasets/movielens/}}, ML-1M, and Netflix\footnote{\url{http://netflixprize.com}}. After binarizing the feedback, we first split the dataset into training, validation, and test sets with a ratio of $70/10/20$. Then, we randomly sample $10\%$ of the interactions from the training set to form the forgetting set $\mathcal{D}_f$, leaving the rest of the training data as the retained set $\mathcal{D}_r$.
\begin{table}[ht]
\vspace{-1pt}
\caption{Statistics of three public datasets.}
\label{tab:dataset_stats}
\centering
\begin{tabular}{lccccc}
\toprule
Dataset & Users & Items & Interactions & Avg. Inter.  & Sparsity \\ 
\hline
ML-100K & 944 & 1,683 & 100,000 & 106 & 93.71\% \\
ML-1M & 6,041 & 3,707 & 1,000,209 & 165 & 95.53\% \\
Netflix & 15,060 & 13,352 & 2,545,835 & 245 & 91.83\% \\
\bottomrule
\end{tabular}
\vspace{-10pt}
\end{table}

\subsubsection{Baselines}

We evaluate CEU on two backbone recommenders, BPR~\cite{rendle2012bpr} and LightGCN~\cite{he2020lightgcn}, to cover both matrix-factorization-based and graph-based settings. 
We compare CEU with representative unlearning methods from three categories.

\begin{itemize}
    \item \textbf{Retraining-based methods.} 
    Retrain removes the forgotten data and retrains the model from scratch on the retained set, serving as the gold standard. 
    GraphEraser~\cite{chen2022graph} and RecEraser~\cite{chen2022recommendation} reduce retraining cost by restricting updates to affected graph partitions or data shards.

    \item \textbf{Fine-tuning-based methods.} 
    GNNDelete~\cite{cheng2023gnndelete} and UnlearnRec~\cite{chen2025pre} perform unlearning by updating lightweight components during the unlearning stage, so that the representations influenced by removed interactions can be revised.

    \item \textbf{Influence-based methods.} 
    SCIF~\cite{li2023selective} and IFRU~\cite{zhang2024recommendation} estimate the influence of removed data and apply corrective parameter updates without full retraining.

\end{itemize}
All methods are evaluated under the same data split and forgetting setting for fair comparison.

\subsubsection{Evaluation Metrics}
We evaluate recommendation utility with HR@$K$ and NDCG@$K$ for $k \in \{5,10\}$~\cite{tamm2021quality,wang2013theoretical}, where higher values indicate better retained-data performance. 
We evaluate unlearning completeness with Membership Inference Attack (MIA) accuracy~\cite{shokri2017membership}; values closer to $0.5$ indicate stronger forgetting. 
We evaluate efficiency by measuring the end-to-end latency of unlearning requests. More metric details are provided in Appendix.

\begin{table*}[ht]
\centering
\caption{Results of unlearning completeness evaluated by Membership Inference Attack (MIA). ACC@1 represents accuracy with 1\% forgetting, while ACC@10 represents accuracy with 10\% forgetting. Attacker access: Both the original score $\hat{y}_{ui}$ and the adjusted score $\hat{y}_{ui}^{\prime}$ are available to the attacker.}
\label{tab:RQ2}
\renewcommand{\arraystretch}{1}
\setlength{\tabcolsep}{6pt}
\newcommand{\na}{\textemdash}
\begin{tabular}{llcccccc}
\toprule
\multirow{2}{*}{\textbf{Dataset}} & \multirow{2}{*}{\textbf{Metric}}
& \multicolumn{3}{c}{\textbf{LightGCN}} & \multicolumn{3}{c}{\textbf{BPR}}\\
\cmidrule(lr){3-5}\cmidrule(l){6-8}
& & Origin & Retrain & CEU & Origin & Retrain & CEU\\
\midrule
\multirow{2}{*}{ML-100K}
& ACC@1 & 0.9560 & 0.5342 & 0.7370 & 0.9797 & 0.5380 & 0.8476 \\
& ACC@10 & 0.9581 & 0.6567 & 0.5155 & 0.9801 & 0.6556 & 0.6429 \\
\midrule
\multirow{2}{*}{ML-1M}
& ACC@1 & 0.9678 & 0.5126 & 0.4789 & 0.9619 & 0.5124 & 0.6847 \\
& ACC@10 & 0.9801 & 0.5015 & 0.6429 & 0.9626 & 0.6443 & 0.5083 \\
\midrule
\multirow{2}{*}{Netflix}
& ACC@1 & 0.9885 & 0.6538 & 0.5581 & 0.9819 & 0.6538 & 0.5581 \\
& ACC@10 & 0.9955 & 0.7059 & 0.5000 & 0.9920 & 0.6029 & 0.5089\\
\bottomrule
\end{tabular}
\end{table*}
\begin{table*}[t]
\centering
\vspace{-8pt}
\caption{Precomputation cost and cache footprint of CCI (FP32).
We report dataset-level totals and per-user averages in parentheses.}
\label{tab:cci_cache}
\small
\begin{tabular}{lcccc}
\toprule
Dataset & Forgotten Users & Time (total) & Cache per-User & Cache Size (Total) \\
\midrule
ML-100K & 9  & 3.10 s/user (28.30s) & 22.45 KiB & 202.05 KiB \\
ML-1M   & 60 & 10.10 s/user (600.64s)  & 22.75 KiB & 1365.00 KiB \\
Netflix & 30 & 8.40 s/user (240.12s)   & 21.88 KiB & 656.40 KiB \\
\bottomrule
\vspace{-15pt}
\end{tabular}
\end{table*}

\subsection{Performance Comparison (RQ1)}
We evaluate utility preservation under a 1\% forgetting ratio on two datasets with BPR and LightGCN, reporting HR@K and NDCG@K ($K\in\{10,20\}$) as in Table~\ref{tab:RQ1}.
Here, Origin denotes the original model before unlearning, while Retrain denotes the gold-standard model retrained on the retained data; both are reported only as reference results and are not included in the comparison among practical unlearning methods.
CEU consistently surpasses approximate baselines, nearing original performance across all settings. On ML-1M with LightGCN, CEU achieves HR@10 = 0.7584 and NDCG@10 = 0.2671, with minimal drops from origin's 0.7603 and 0.2682 (0.25\% and 0.41\% reductions), outperforming SCIF (HR@10 = 0.6060) by 20.29\% and NDCG@10 by 21.14\%. Retraining-based methods like RecEraser (HR@10 = 0.6003) and GraphEraser (HR@10 = 0.5336) lag significantly (21.00\% and 29.64\% below CEU on ML-1M LightGCN) due to data fragmentation. 
Similarly, fine-tuning-based methods (e.g., GNNDelete) also suffer from utility degradation, 
while CEU's targeted intervention preserves utility effectively. This trend holds across datasets.
These results confirm that CEU successfully preserves model utility (G3), avoiding the significant collateral damage to non-target data that plagues other approximate and SISA-based methods.
Additional results in the Appendix further show that, as the forgetting ratio increases, baseline methods exhibit clear degradation on the retained set, while CEU remains much more stable.






\subsection{Unlearning Completeness (RQ2)}
Following prior work~\cite{li2024making,li2023selective}, we employ a membership inference attack to assess the extent to which the forgotten data has been removed.
For this attack, the adversary is assumed to have access to both the model’s raw prediction logits ($\hat{y}_{ui}$) and the final post-processed recommendation scores ($\hat{y}_{ui}^{\prime}$). 
All reported attack results are based on this full-knowledge white-box threat model.
We report the attacker's accuracy (ACC) at 1\% and 10\% forgetting ratios in Table~\ref{tab:RQ2}. 
To rigorously evaluate absolute completeness, we directly benchmark CEU against the unmitigated Origin model and the ideal Retrain standard. 
CEU significantly reduces MIA accuracy, approaching Retrain’s ideal forgetting performance while outperforming the unmitigated Origin model. For instance, on ML-1M with LightGCN, CEU achieves ACC@1 = 0.4789 and ACC@10 = 0.6429, a substantial drop from Origin’s 0.9678 and 0.9801 (50.53\% and 34.42\% reductions), and closely aligns with Retrain’s 0.5126 and 0.5015. On Netflix with LightGCN, CEU’s ACC@10 = 0.5000 nearly matches random guessing, compared to Origin’s 0.9955, highlighting its superior forgetting efficacy.
These results confirm that CEU effectively erases forgotten data, fulfilling the Unlearning Completeness (G1) objective.
\begin{table}[t]
\centering
\caption{
  Unlearning latency (s). Bold and underlined values indicate the
  best and second-best results, respectively; Speedup is measured
  against the second-best method.
}
\label{tab:RQ3}
\footnotesize
\setlength{\tabcolsep}{1.2pt}
\renewcommand{\arraystretch}{1.05}

\begin{tabular}{@{}lcccccc@{}}
\toprule
\multirow{2}{*}{\textbf{Method}}
& \multicolumn{2}{c}{\textbf{ML-100K}}
& \multicolumn{2}{c}{\textbf{ML-1M}}
& \multicolumn{2}{c}{\textbf{Netflix}} \\
\cmidrule(lr){2-3}
\cmidrule(lr){4-5}
\cmidrule(l){6-7}
& BPR & LightGCN
& BPR & LightGCN
& BPR & LightGCN \\
\midrule
Retrain
& 248.37 & 141.92
& 1935.41 & 4209.05
& 527.66 & 1019.13 \\

GraphEraser
& 26.08 & \underline{17.93}
& 302.11 & 269.74
& 310.06 & 640.82 \\

RecEraser
& 29.31 & 35.19
& 386.57 & 275.02
& 404.68 & 880.21 \\

SCIF
& \underline{18.36} & 18.50
& 66.72 & 64.09
& 299.88 & 177.14 \\

IFRU
& 55.07 & 57.63
& 78.40 & 90.55
& \underline{104.12} & \underline{117.95} \\

GNNDelete
& 32.94 & 35.39
& 68.27 & 67.79
& 152.38 & 177.17 \\

UnlearnRec
& 19.21 & 20.03
& \underline{51.68} & \underline{59.32}
& 113.58 & 123.16 \\

\textbf{CEU}
& \textbf{0.36} & \textbf{0.34}
& \textbf{4.50} & \textbf{6.37}
& \textbf{2.81} & \textbf{2.89} \\
\midrule
\textbf{Speedup}
& 51.00$\times$ & 52.74$\times$
& 11.48$\times$ & 9.31$\times$
& 37.05$\times$ & 40.81$\times$ \\
\bottomrule
\end{tabular}
\end{table}
\begin{figure*}[ht]
  \centering
  \includegraphics[width=.85\textwidth]{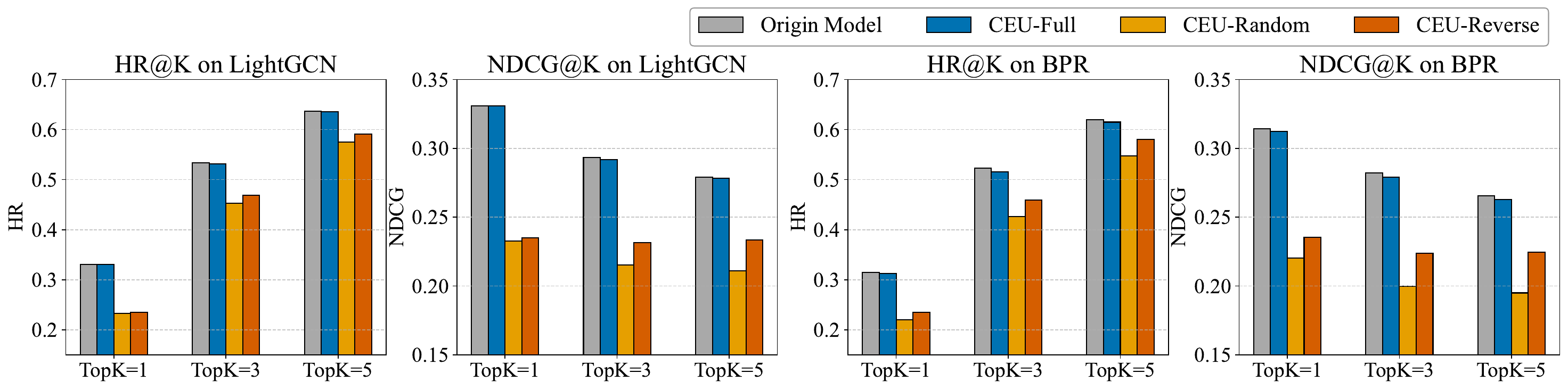}
  \caption{Comparison of CEU variations on ML-1M with LightGCN and BPR backbones. CEU-Full preserves HR@20 and NDCG@20 nearly identical to the original model, whereas CEU-Random and CEU-Reverse cause substantial performance degradation.}
  \label{fig:ablation_utility_bars}
\end{figure*}
\begin{figure*}[!ht]
  \centering
  \vspace{-5pt}
  \includegraphics[width=0.7\textwidth]{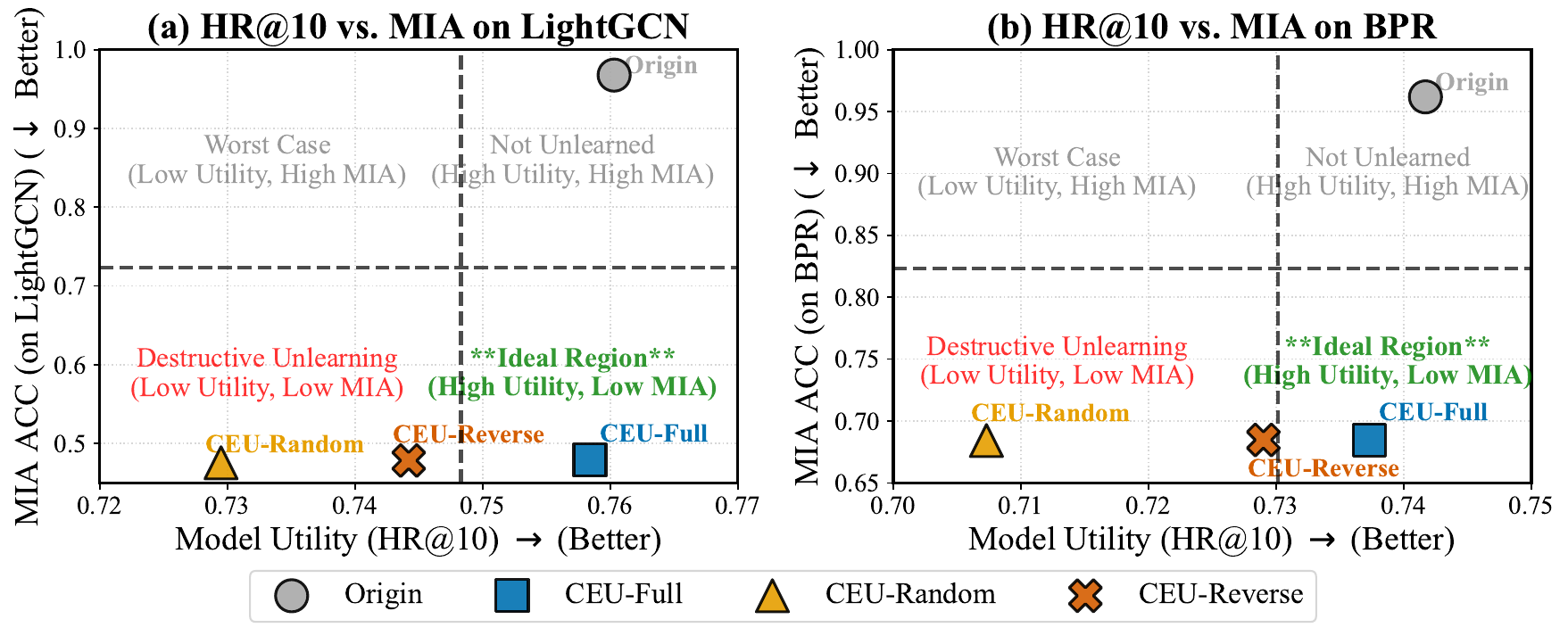}
  \caption{Trade-off between recommendation utility and unlearning effectiveness on ML-1M under LightGCN and BPR. The bottom-right region is the ideal zone with high utility and low MIA accuracy. CEU-Full lies within this ideal zone, demonstrating precise, non-destructive forgetting, whereas CEU-Random and CEU-Reverse significantly degrade utility and lower MIA accuracy.}
  \vspace{-5pt}
  \label{fig:ablation_tradeoff}
\end{figure*}
\subsection{Unlearning Efficiency (RQ3)}
We evaluate efficiency from two complementary perspectives: online unlearning latency and offline CCI precomputation cost. Table~\ref{tab:RQ3} reports the end-to-end latency for a single unlearning request. CEU consistently achieves the lowest latency ($0.34$--$6.37$\,s) across all datasets and backbones, outperforming the second-best baselines by $10.06\times$ to $55.64\times$. In contrast, full retraining requires $141.92$--$4209.05$\,s, and other baselines remain one to three orders of magnitude slower. This highlights CEU's real-time efficiency, achieved entirely through parameter-free score adjustments.

To ensure transparency, Table~\ref{tab:cci_cache} details the optional offline CCI precomputation cost. Using a single NVIDIA RTX 4090 GPU, we cache the Top-50 candidates (with up to five causal factors per item) for each user. The per-user computation time is modest ($3.10$--$10.10$\,s), yielding a compact cache footprint of approximately $22$\,KiB that can be amortized across future requests. 

For large-scale deployments, CEU can also operate on demand. Executing the CCI step only when a user actually submits an unlearning request reduces the need for full precomputation and storage while preserving the same online unlearning mechanism. Overall, these results confirm that CEU meets the efficiency goal (G2) and supports low-latency unlearning in practice.

\subsection{Effectiveness of CCI (RQ4)}
To validate the Causal Contribution Identifier (CCI) module, we conducted an ablation study on the ML-1M dataset comparing our full method (\textbf{CEU-Full}) against two non-causal variants: \textbf{CEU-Random} (applying random penalties) and \textbf{CEU-Reverse} (penalizing causally unrelated items). As shown in Fig.~\ref{fig:ablation_utility_bars} and Fig.~\ref{fig:ablation_tradeoff}, \textbf{CEU-Full} almost perfectly preserves the original model's utility (e.g., LightGCN's NDCG@10 drops by a mere 0.4\%, from 0.2682 to 0.2671), whereas the non-causal variants trigger a catastrophic performance collapse (e.g., \textbf{CEU-Random}'s NDCG@10 plummets by 19.4\% on LightGCN and 20.8\% on BPR). Furthermore, while these non-causal variants also reduce MIA accuracy to $\approx$0.48, this low MIA score is a false positive due to the complete collapse of the model's ranking ability. Ultimately, these results demonstrate that the CCI module is indispensable; it provides the precise causal guidance needed to achieve precise unlearning without sacrificing overall model utility.

\section{Conclusion}
In this paper, we addressed recommendation unlearning under the strict asymmetric access constraints of MLaaS architectures, where parameter updates are prohibitive. To tackle the lack of interpretability and side-effect propagation in existing methods, we proposed CEU, a novel parameter-free “Explain, then Forget” framework. CEU first identifies minimal causal support sets via CCI and then performs localized score adjustment via URA to remove the influence of forgotten interactions without modifying model parameters. Extensive experiments across three benchmarks verify that CEU preserves recommendation utility close to that of the original model while significantly outperforming existing unlearning baselines in both forgetting quality and efficiency. The ablation study further confirms that causal guidance is essential for avoiding destructive side effects and maintaining model fidelity. Overall, CEU offers a practical solution for efficient and reliable recommendation unlearning in API-only MLaaS environments.

\bibliographystyle{elsarticle-num}
\bibliography{conference}

\appendix





\section*{Stability under Larger Forget Ratios}
\label{app:forget_ratio_stability}

\begin{figure}[ht]
  \centering
  \includegraphics[width=0.48\textwidth]{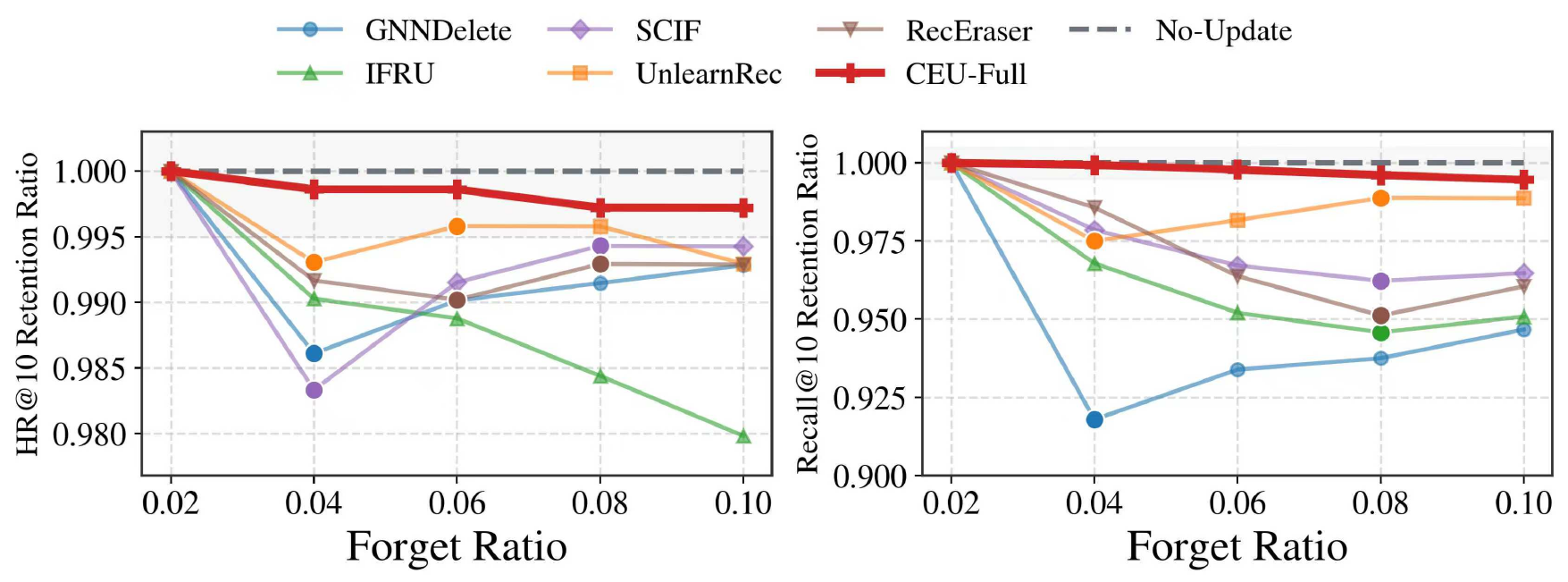}
  \caption{Retained-set stability under increasing forget ratios. The left panel shows the HR@10 retention ratio, and the right panel shows the Recall@10 retention ratio, both normalized by the value at the $2\%$ forget ratio. Higher values indicate better retained-set stability under stronger forgetting.}
  \label{fig:appendix_forget_ratio}
  \vspace{-5pt}
\end{figure}

To further examine the stability of different unlearning methods, we evaluate retained-set performance under increasing forget ratios. Specifically, we vary the forget ratio from $2\%$ to $10\%$ and normalize the retained-set performance of each method by its value at the $2\%$ forget ratio. Thus, the vertical axis reports the relative retention ratio. A value closer to $1.0$ indicates better preservation of retained-set utility as more interactions are forgotten.
We note that this experiment is not a strict simulation of sequential request streams. Instead, it serves as a stress test of retained-set stability under stronger forgetting. As shown in Figure~\ref{fig:appendix_forget_ratio}, most baseline methods show a clear downward trend as the forget ratio increases, although the degree of degradation varies across methods. In contrast, CEU remains much more stable across all settings and stays close to the no-update reference.

This result is consistent with the design of CEU. Since CEU performs unlearning via score adjustment rather than repeated parameter updates, it is less sensitive to utility degradation during repeated or large-scale forgetting. Therefore, this experiment provides supplementary evidence that the parameter-free design of CEU helps preserve retained-set utility as the forgetting intensity increases.

\section*{Pseudo Algorithms of CEU}
\label{app:algorithms}

This appendix summarizes the main CEU procedure in algorithmic form. CEU consists of two stages. First, the \textbf{Causal Contribution Identifier (CCI)} explains a recommendation by identifying the historical interactions that causally support it and by estimating their contribution strengths. 
Second, the \textbf{Unlearning-driven Rank Adjuster (URA)} uses these causal signals to remove the influence of a forgotten interaction by adjusting scores, without modifying model parameters.

\begin{algorithm}[t]
\caption{Causal Contribution Identifier (CCI)}
\label{alg:cci}
\small
\KwIn{
pretrained recommender $f(\cdot \mid \theta)$; user $u$; target item $i$; user history $\mathcal{H}_u$; top-$K$ cutoff $K$; margin $m$; sparsity weight $\lambda$; threshold $\delta$
}
\KwOut{
causal explanation $E_{ui} = (\mathcal{S}_{ui}^{*}, w_{ui})$
}

Initialize mask vector $\mathbf{P}_u \leftarrow \mathbf{1}$\;
Obtain the original top-$K$ list for user $u$ and let $i_{K+1}$ be the item ranked at position $K+1$\;

\Repeat{convergence}{
Construct the masked history $\mathcal{H}_u \odot \mathbf{P}_u$\;
Compute the flip loss
\[
\mathcal{L}_{\mathrm{flip}}
=
\max \big(0,\; m + f(u^{(\mathbf{P}_u)}, i \mid \theta) - f(u, i_{K+1} \mid \theta)\big)
\]
Update $\mathbf{P}_u$ by minimizing
\[
\mathcal{L}_{\mathrm{flip}} + \lambda \|\mathbf{1}-\mathbf{P}_u\|_1
\]
}

Compute the causal support set
\[
\mathcal{S}_{ui}^{*} = \{ h \in \mathcal{H}_u \mid P_{u,h} < 1-\delta \}
\]
For each $h \in \mathcal{S}_{ui}^{*}$, compute its contribution strength
\[
w_{ui}(h) = 1 - P_{u,h}
\]
\Return{$E_{ui} = (\mathcal{S}_{ui}^{*}, w_{ui})$}\;
\end{algorithm}

\begin{algorithm}[t]
\caption{Unlearning-driven Rank Adjuster (URA)}
\label{alg:ura}
\small
\KwIn{
pretrained recommender $f(\cdot \mid \theta)$; user $u$; forgotten interaction $h_f=(u,i_f)$; current recommendation list $\mathcal{L}_u$; cached explanations $\{E_{ui}\}_{i \in \mathcal{L}_u}$; penalty factor $\alpha$
}
\KwOut{
updated recommendation list $\mathcal{L}'_u$
}

\ForEach{$i \in \mathcal{L}_u$}{
Compute the original score $\hat{y}_{ui} = f(u,i \mid \theta)$\;
Retrieve $E_{ui} = (\mathcal{S}_{ui}^{*}, w_{ui})$\;
\eIf{$h_f \in \mathcal{S}_{ui}^{*}$}{
Adjust the score by
\[
\hat{y}'_{ui} = \hat{y}_{ui} - \alpha \cdot \hat{y}_{ui} \cdot w_{ui}(h_f)
\]
}{
Set $\hat{y}'_{ui} = \hat{y}_{ui}$\;
}
}
Re-rank items in $\mathcal{L}_u$ according to $\{\hat{y}'_{ui}\}$\;
Obtain the updated recommendation list $\mathcal{L}'_u$\;
\Return{$\mathcal{L}'_u$}\;
\end{algorithm}

In practice, CCI can be run offline to pre-compute explanations for candidate items and cache them for later use. 
When an unlearning request arrives, URA reads the cached explanations and updates the ranking scores online. 
This design makes CEU suitable for low-latency unlearning in API-only environments.

\section*{Experiment Setup Details}

\subsection{Evaluation Metrics}

We evaluate recommendation utility using two standard ranking metrics: Hit Ratio (HR@$K$) and Normalized Discounted Cumulative Gain (NDCG@$K$)~\cite{tamm2021quality,wang2013theoretical}. 
Higher values indicate better recommendation quality on the retained data after unlearning. We evaluate unlearning completeness using Membership Inference Attack (MIA) accuracy~\cite{shokri2017membership}. This metric measures how well an attacker can distinguish between forgotten and retained interactions. An MIA accuracy closer to $0.5$ means the forgotten interactions are harder to identify, indicating better forgetting. We evaluate unlearning efficiency by recording the end-to-end latency of unlearning requests. Lower latency indicates stronger efficiency.

\subsection{Implementation Details}

All experiments are conducted on a server with one NVIDIA RTX 4090 GPU. We fix the random seed to 2020 for all runs. Unless otherwise stated, all models are trained with Adam for 300 epochs, using a learning rate of 0.001 and a training batch size of 2048. We evaluate the model on the validation set after each epoch and apply early stopping with a patience of 10, where Recall@10 is used as the validation metric. For evaluation, we adopt the full-ranking setting and report Recall, NDCG, and Hit at $K \in {10,20}$, with an evaluation batch size of 4096.

For CEU, the CCI module pre-computes causal explanations for the top 50 candidate items of each user. For each candidate item, we retain at most 5 causal interactions as its explanatory support set. These cached explanations are then used by URA during the unlearning stage to perform score adjustment without updating model parameters.

\end{document}